\documentclass[conference]{IEEEtran}
\IEEEoverridecommandlockouts
\usepackage{cite}
\usepackage[numbers]{natbib}
\usepackage{amsmath,amssymb,amsfonts}
\usepackage{algorithmic}
\usepackage{graphicx}
\usepackage{textcomp}
\usepackage{xcolor}
\usepackage{subfigure}
\usepackage{dirtytalk}
\usepackage{comment}
\usepackage{adjustbox}
\usepackage{balance}
\usepackage{dblfloatfix} 
\usepackage[hyphens]{url}
\makeatletter
\newcommand{\linebreakand}{%
  \end{@IEEEauthorhalign}
  \hfill\mbox{}\par
  \mbox{}\hfill\begin{@IEEEauthorhalign}
}
\makeatother

\def\BibTeX{{\rm B\kern-.05em{\sc i\kern-.025em b}\kern-.08em
    T\kern-.1667em\lower.7ex\hbox{E}\kern-.125emX}}
\begin{document}
\bstctlcite{IEEEexample:BSTcontrol}

\title{The Usability and Trustworthiness of Medical Eye Images\\
}

\author{\IEEEauthorblockN{Daniel Diethei}
\IEEEauthorblockA{
\textit{University of Bremen}\\
Bremen, Germany \\
diethei@uni-bremen.de}
\and
\IEEEauthorblockN{Ashley Colley}
\IEEEauthorblockA{
\textit{University of Lapland}\\
Rovaniemi, Finland \\
ashleycolley@gmail.com}
\and
\IEEEauthorblockN{Lisa Dannenberg}
\IEEEauthorblockA{
\textit{University of Bremen}\\
Bremen, Germany \\
lisadannenberg2@gmail.com}
\linebreakand
\IEEEauthorblockN{Muhammad Fawad Jawaid Malik}
\IEEEauthorblockA{
\textit{University of Bremen}\\
Bremen, Germany \\
malikm@uni-bremen.de}
\and
\IEEEauthorblockN{Johannes Schöning}
\IEEEauthorblockA{
\textit{University of Bremen}\\
Bremen, Germany \\
schoening@uni-bremen.de}
}

\maketitle

\begin{abstract}
The majority of blindness is preventable, and is located in developing countries. While mHealth applications for retinal imaging in combination with affordable smartphone lens adaptors are a step towards better eye care access, the expert knowledge and additional hardware needed are often unavailable in developing countries. Eye screening apps without lens adaptors exist, but we do not know much about the experience of guiding users to take medical eye images. Additionally, when an AI based diagnosis is provided, trust plays an important role in ensuring in the adoption. This work addresses factors that impact the usability and trustworthiness dimensions of mHealth applications. We present the design, development and evaluation of \textit{EyeGuide}, a mobile app that assists users in taking medical eye images using only their smartphone camera. In a study (n=28) we observed that users of an interactive tutorial captured images faster compared to audible tone based guidance. In a second study (n=40) we found out that providing disease-specific background information was the most effective factor to increase trustworthiness in the AI based diagnosis. Application areas of \textit{EyeGuide} are AI based disease detection and telemedicine examinations.
\end{abstract}


\section{Introduction \& Motivation}
According to the World Health Organization (WHO), 65.2 million people have cataract, a cloudiness in the lens of the eye, leading to increasingly blurred vision~\cite{WHO2019vision}. Of the visually impaired and blind, 90\% live in developing countries and 80\% of visual impairments are preventable ~\cite{WHO2013}. In many countries, the poorest households are more likely to have access to mobile phones than to toilets or clean water~\cite{world2016world}. With this background, mobile health (mHealth) technologies that can support the early detection of causes of preventable blindness should be explored with high priority.
Retinal screening programs for common eye diseases can provide early detection of chronic eye diseases, but come at a high cost, e.g., a stationary slit lamp camera for traditional ophthalmology costs about \$15.000~\cite{russo2015comparison}. One step to increase the affordability of retinal screening is to use a shared community smartphone fitted with a lens adaptor, e.g. Peek Retina (\$240) or D-Eye (\$435). Such solutions can be applied by clinicians and nurses in areas with limited healthcare facilities and have been shown to perform well in disease detection.~\cite{russo2015comparison}

Whilst the increasing availability of smartphones has resulted in a growing market for mHealth applications, e.g. in the areas of respiratory diseases~\cite{Lamonaca2015} and dermatology~\cite{MySkinSelfie}, there is little evidence of positive clinical outcomes~\cite{tomlinson2013}. Studies have suggested that users require consumer-friendly devices and apps that are self-reinforcing and enjoyable to use~\cite{steinhubl2015emerging}. Reporting on experiences in the UK's National Health Service, ~\cite{greenhalgh2004spread} highlight ease of use of mHealth apps as a deciding factor in their adoption. A further barrier to the adoption of mHealth services is their perceived trustworthiness, e.g. in the competence of the provider, privacy or security of the service~\cite{angst2009adoption,guo2016privacy,KHATUN2015determinants}.

\begin{figure}
\centering
  \includegraphics[width=\columnwidth]{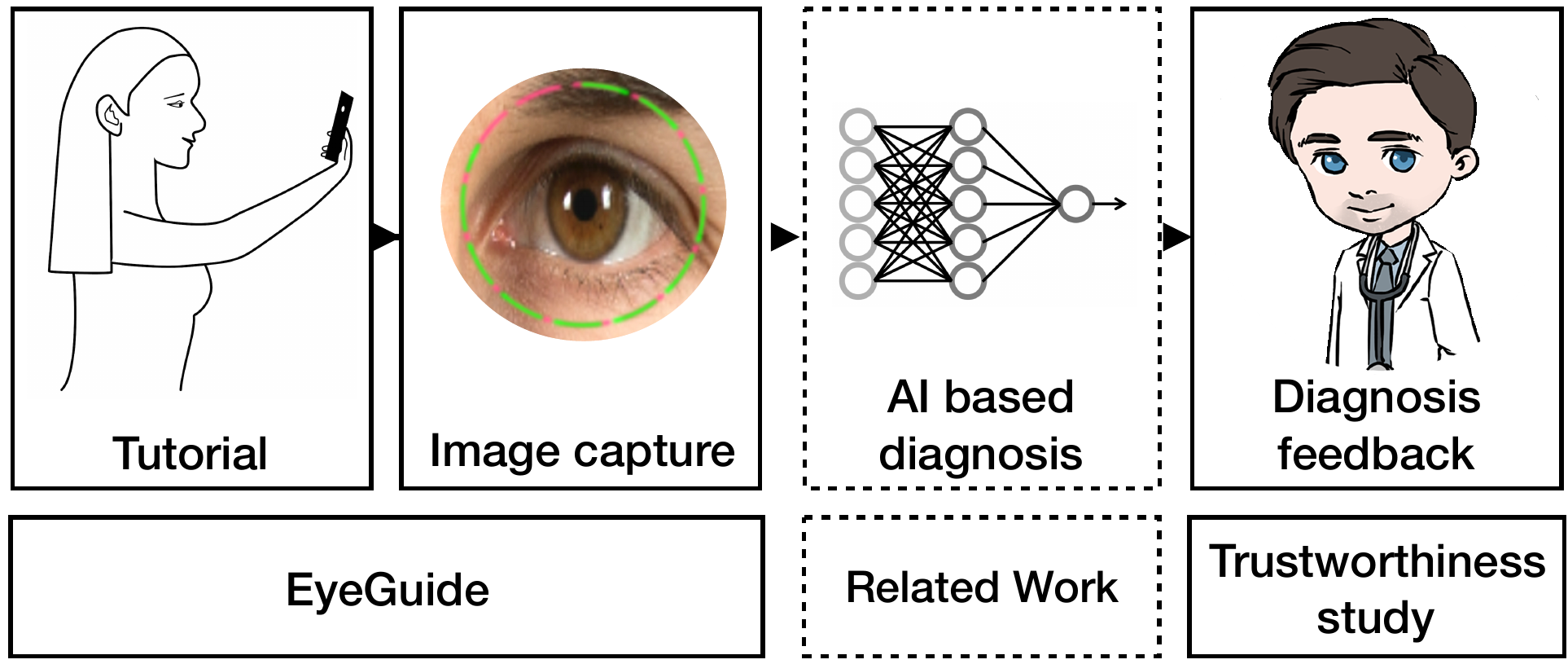}
  \caption{Smartphone based eye disease diagnosis app. The scope of this paper covers (1) the tutorial and image capture phases and (2) the diagnosis feedback, focusing on the trustworthiness of the diagnosis. The AI based diagnosis functionality is not in the scope of this paper. }~\label{fig:figure1}
\end{figure}


To address the issues mentioned above, we developed a user-friendly smartphone app for eye disease detection. In this paper, we focus specifically on the phases of image capture and the presentation of the diagnosis to the user (Figure \ref{fig:figure1}). To account for a wide range of individual's living circumstances, we target a solution enabling self-taking of medical eye images, and focus on the use of the smartphone's higher quality rear camera. Whilst there is much prior work on automated diagnosis of eye disease from images, e.g. using deep learning~\cite{xu2017deep,wei2019deep,lam2018retinal,padhy2019artificial}, the effectiveness of this is limited if the source images for analysis are of poor quality.

The contribution of this work is twofold as we connect the fields of usability and trustworthiness in the context of mHealth. As suggested by others~\cite{roy2001impact,egger2000trust,kim1998designing}, the usability of mHealth applications impacts trustworthiness, i.e. users find applications that consider usability principles to be more trustworthy. Therefore,  we describe the design, development and evaluation of the multi-modal \textit{EyeGuide} app, that guides users in the capture of eye images suitable for detection of eye diseases such as cataracts. To support the task, the app includes a tutorial phase and uses auditory and visual guidance to direct the user to position the camera correctly. As a second contribution, we present a study exploring different approaches to increase the  perceived trustworthiness of the eye disease diagnosis app.

From our evaluation, we report that the interactive tutorial had advantages in terms of usability of  quality of taken images compared to the auditory guidance. A review of the taken images by an ophthalmologist indicated that the images taken with our app are suitable for diagnosing diseases in the conjunctiva and cornea. The most effective way to build trustworthiness in the app was through increasing users' knowledge level about the disease.

\section{Related Work}
Our work in the area of intelligent user interfaces builds on three areas of prior work on (1) smartphone based mHealth for eye diseases, (2) guiding image capture, and (3) trustworthiness in mHealth. Firstly, we introduce existing mHealth applications addressing eye-related diseases. We then present works with different approaches to guide users when taking photographs, e.g. ensuring the subject is correctly located in the frame. Finally, we give a short overview of issues surrounding the perceived trustworthiness of mHealth applications. 

\subsection{Smartphone-Based Eye Disease Diagnosis}
Prior work has presented smartphone apps to diagnose disease based on eye-images, e.g. for cataract detection~\cite{rana2017,pathak2016robust}, to identify high cholesterol levels ~\cite{Alhasawi18,kumar2016noninvasive}, to diagnose concussions \cite{mariakakis2017pupil} and for glaucoma screening~\cite{kumar2016noninvasive,guo2018yanbao}. ~\citet{akil2019detection} present a meta paper, investigating the image quality and diagnosis performance achieved in eight prior works using smartphones equipped with additional lenses for retinal examination. Most recently deep learning has been presented as an approach to identify eye diseases~\cite{xu2017deep,wei2019deep,lam2018retinal,padhy2019artificial}. For example, ~\citet{wei2019deep} presented a deep learning based smartphone app to identify retinal abnormalities. The system gives simple real-time textual feedback `normal / disease detected', requires the use of an additional D-Eye lens fitted to the smartphone and cannot be self-administered. ~\citet{kim2018widefield} developed and evaluated an automated smartphone-based system for retinal disease screening using a 3D printed housing for acquiring a series of seven images stitched together to a widefield retinal montage. ~\citet{Munsoneaax6363} developed a free smartphone app able to identify `white eye' and subsequent eye disorders based on casual photographs of small children. The authors reported the app correctly diagnosed 80\% of children with eye disorders. EyeGuide builds upon work by \citet{Diethei:2018} who explored an agent-based tutorial for eye image guidance. In this paper, we present a study to assess the usability of auditory guidance and an interactive tutorial to take eye images without the need of additional lenses or other hardware.

\subsection{Guiding Image Capture} 
Ensuring correctly positioned and high quality images of the eye is a critical element in the performance of the following diagnosis phases, either manually by a doctor or through AI based solutions. As the target of our approach is towards the creation of a self-administered eye examination tool, using the device's higher quality rear camera, solutions to guide the user in image capture are needed. Similar challenges have been addressed in mHealth apps for dermatology. For example, in the Skinvision app~\cite{Skinvision} the user ``waves" the smartphone over the skin lesion to be investigated, and the app automatically selects when to take the image for analysis. This unguided approach could potentially be improved on with audio feedback~\cite{diethei2020medselfies}, e.g., similar to that used for navigation~\cite{holland2002audiogps}. A summary of approaches to integration non-speech sounds to visual interfaces has been presented by ~\cite{absar2008}, suggesting the use of earcons~\cite{Blattner:1989,heller2018navigatone}, sounds with dynamic timber, pitch and rhythm, as suitable for localization tasks. To guide users to take better portrait photos with smartphones, ~\cite{lo2013yes} demonstrated a solution using verbal guidance e.g. \say{please move to the left}, to which ~\cite{balata2015blindcamera} added vibration modality.

\subsection{Trustworthiness}
The importance of the user interface in perceived trustworthiness has been reported by several works, e.g., ~\cite{egger2000trust,kim1998designing,roy2001impact}. ~\citet{kim1998designing} have shown that the user interface design factors impact customer confidence. \citet{roy2001impact} found out that the usability factors ease of navigation, ease of learning, perception and support were associated with trustworthiness. In the field of eHealth, ~\citet{vo2019patients} highlighted trustworthiness, appropriateness, personalization, and accessibility as the main weaknesses in current mHealth apps. \citet{fruhling2006influence} developed a usability model for the consumer's perception of trust in eHealth services, noting that websites with a higher usability were likely to be viewed as more trustworthy. Going beyond usability, others have highlighted the role of visual aesthetics in the design of mHealth apps in their perceived credibility~\cite{oyibo2018drives}. However, mHealth app designers do not have free hands, the requirements for apps to meet regulatory requirements can impact the design, and consequentially affect the perceived trustworthiness the app~\cite{chatzipavlou2016recommended}.

\begin{figure*}
\centering
  \includegraphics[width=.8\textwidth]{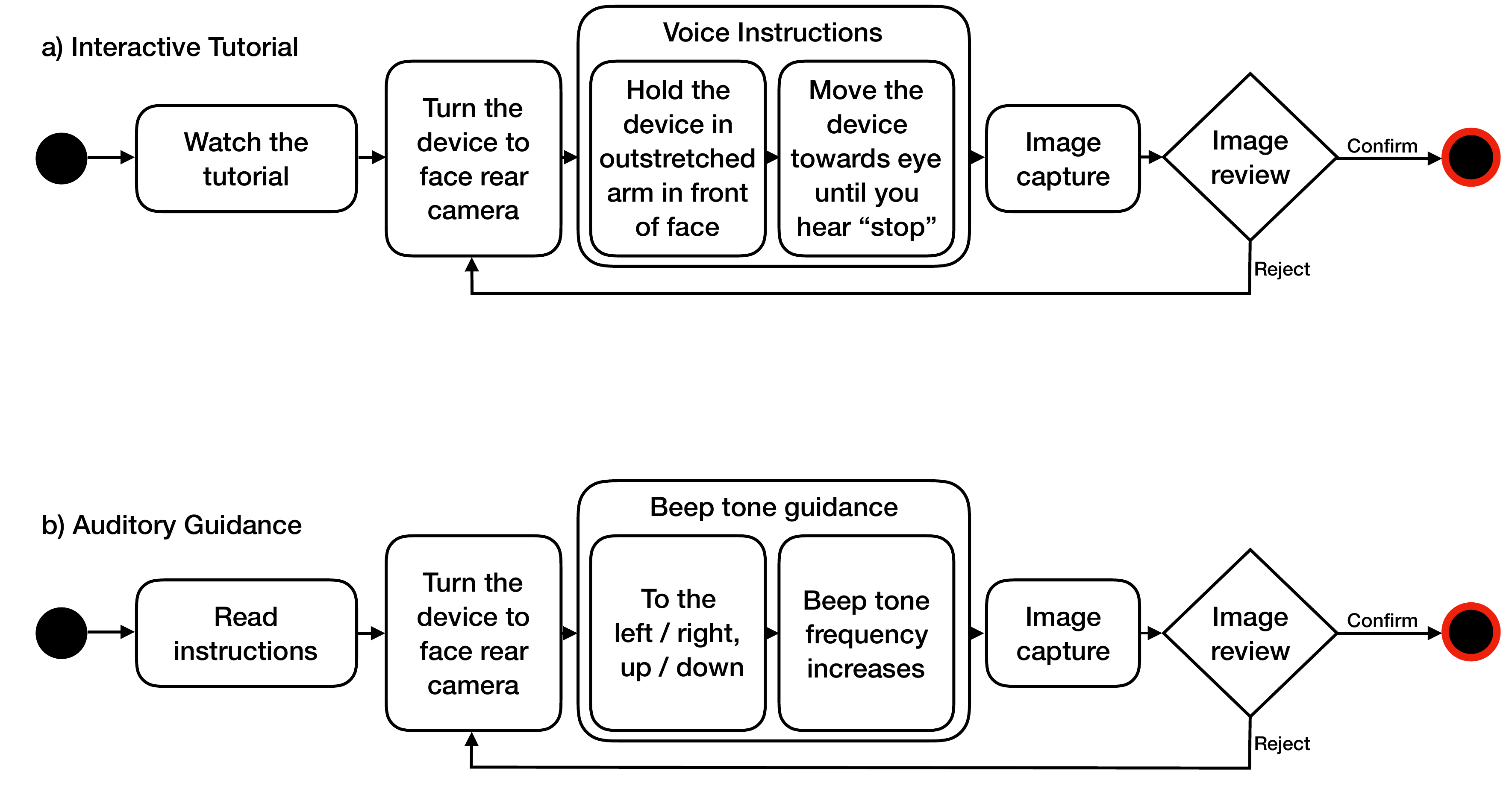}
  \caption{Flow diagram of the two study conditions. Participants in the interactive tutorial group first watched a tutorial demonstrating the procedure to take eye images and then followed voice instructions. In the auditory guidance group, participants only read brief instructions to take an image of the iris and were then supported through voice and beep tone guidance.}~\label{fig:study_conditions}
\end{figure*}

\subsection{Summary of Prior Work}
From the related work we learn that, whilst much research effort has been spent addressing automated eye-disease diagnosis from images, little research has addressed the activities preceding or following this in an operational mHealth solution. With a focus on these phases we, (1) demonstrate and evaluate two solutions using audio signals and an interactive tutorial to guide users to take optimal medical eye images (2) present design alternatives aiming to improve the perceived trustworthiness of mHealth smartphone apps. In particular our approach follows Akter et al.'s ~\cite{akter2011trustworthiness} dimensions of trustworthiness; ability, benevolence, integrity and predictability.

\section{EyeGuide}
The \textit{EyeGuide} app guides the user to take high quality images of their own eye, using the rear camera of a smartphone. These images are suitable for analysis by a clinician or through an automated AI process. As in this configuration the smartphone's screen is facing away from the user, the user must be guided to position the camera to fully capture the eye.

\subsection{Concept}
To understand the issues related to clinical eye photography on a smartphone, we first instructed four individuals to acquire photos of their iris with their smartphone in a prestudy. The participants used the native camera application with both the front and rear camera. An analysis of the resulting images revealed that lighting and device positioning seemed to be the predominant causes of blurry images or images without a fully visible iris. The front camera did not provide sufficient image quality for our purpose of detecting diseases due to the lack of an autofocus functionality and the lower image resolution. Although the front cameras of the latest high-end smartphones are of higher resolution than those used in the test, considering our target populations in developing countries, the use of older, lower-cost devices with limited front camera resolution is more representative.

Based on previous work by~\citet{Diethei:2018} and the cognitive theory of multimedia learning~\cite{clark2016learning}, two variants were implemented:
\begin{itemize}
\item Interactive tutorial (referred to as IT)
\item Reading instructions followed by voice and audio tone based guidance (referred to as `auditory guidance'; AG)
\end{itemize}

The procedures of the two app variants are illustrated in Figure~\ref{fig:study_conditions}.

\newlength{\width}
\setlength{\width}{0.25\textwidth}
\newlength{\horspace}
\setlength{\horspace}{0.3cm}

\begin{figure}[!b]
\begin{adjustbox}{max width=\columnwidth,center}
\centering
\subfigure[]{\label{fig:step1}\includegraphics[width=\width]{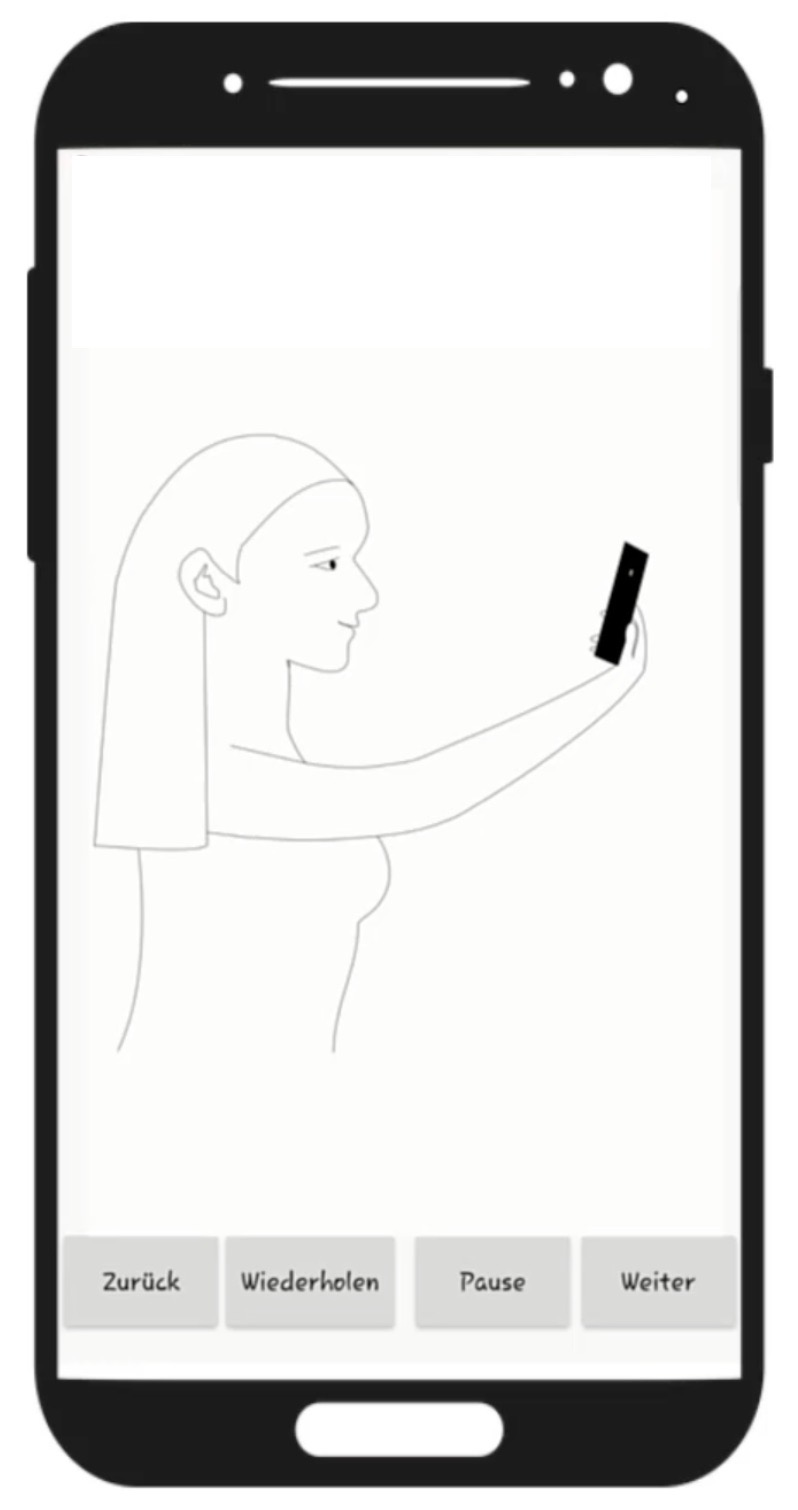}}\hspace{\horspace}
\subfigure[]{\label{fig:step2}\includegraphics[width=\width]{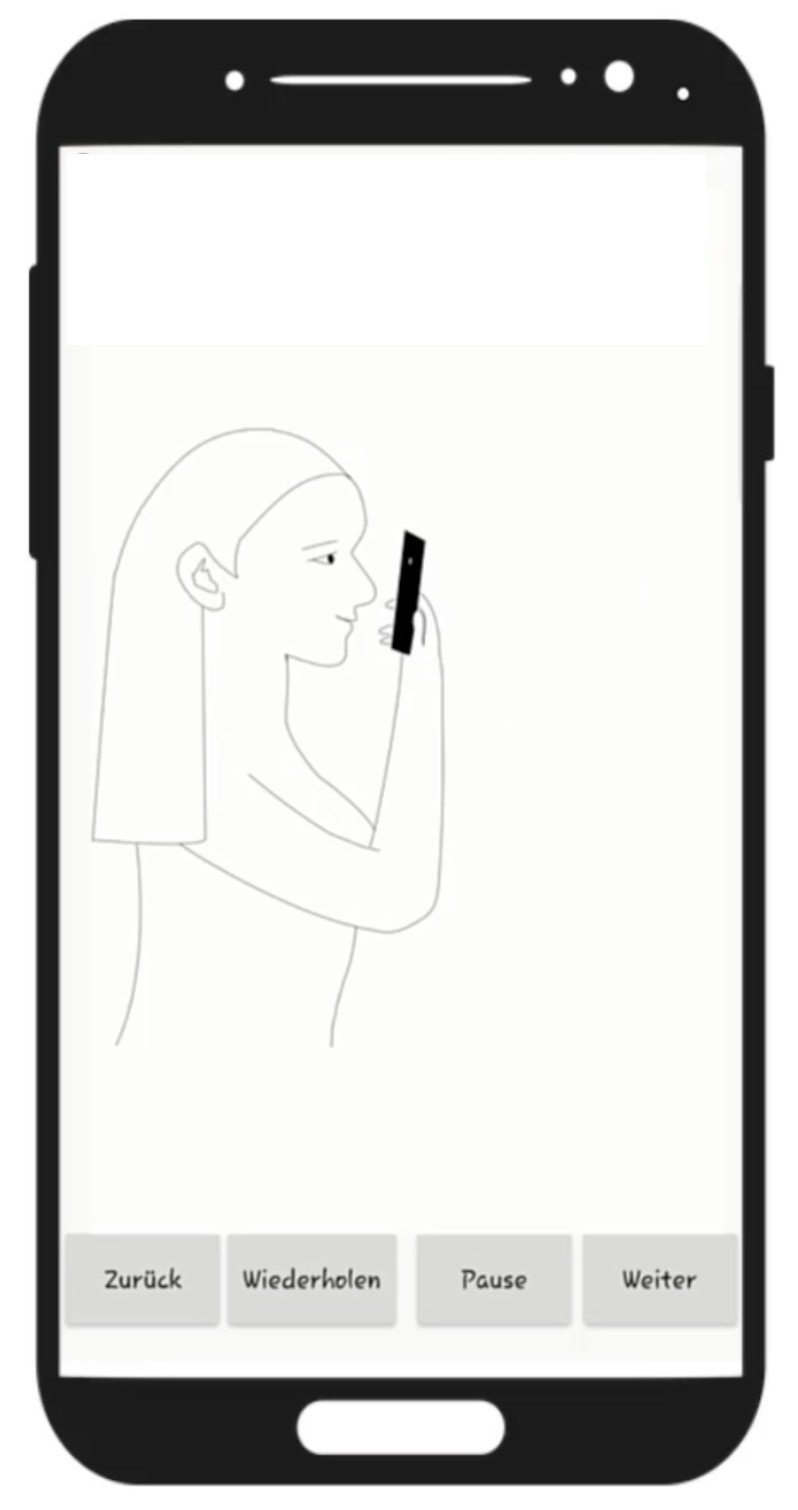}}\hspace{\horspace}
\subfigure[]{\label{fig:step3}\includegraphics[width=\width]{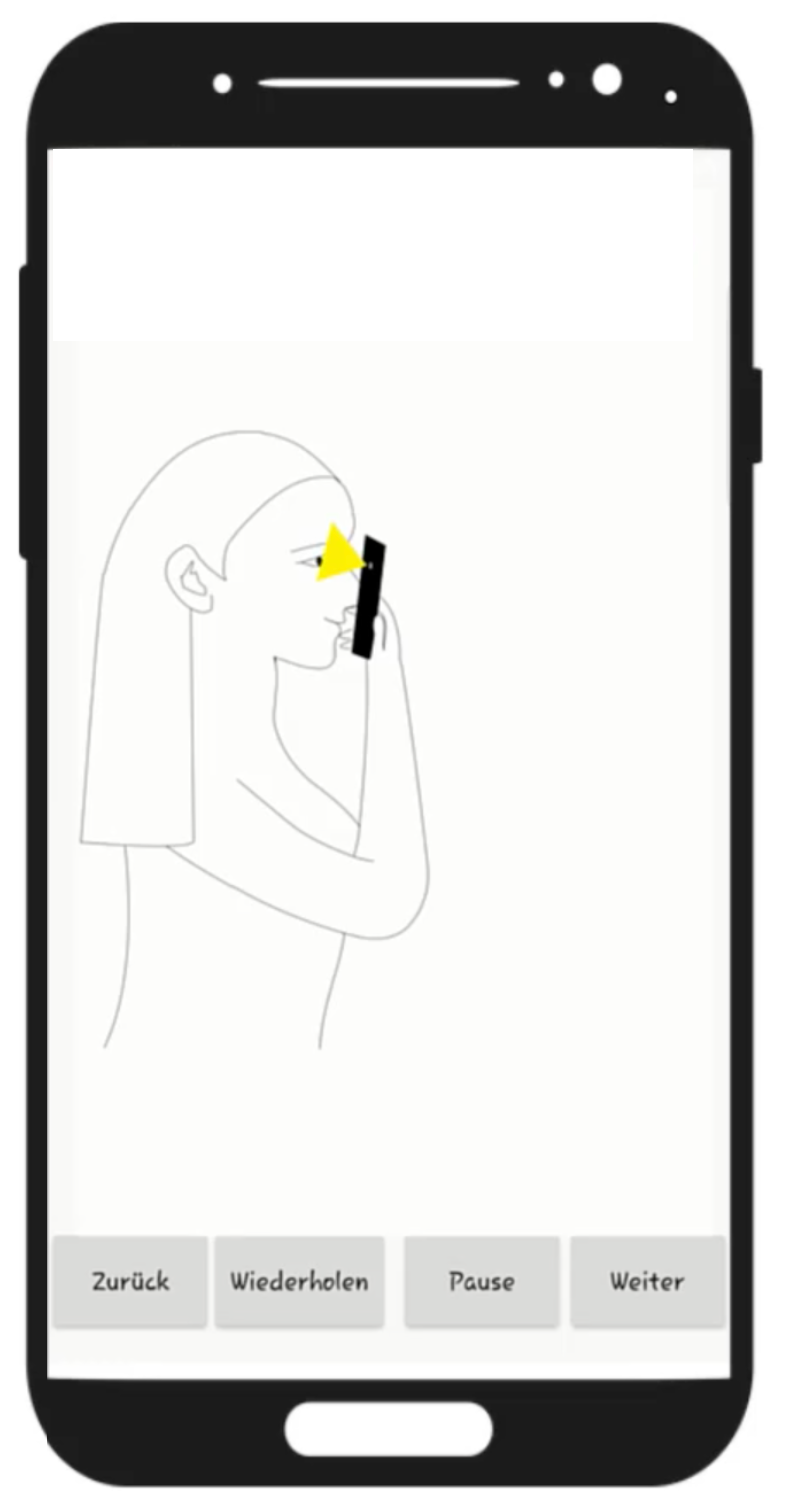}}\hspace{\horspace}
\end{adjustbox}
\caption{Tutorial to take the eye photos. Left (a): Arm outstretched in front of face. Center (b): Moving the phone closer to the eye until the word ``stop" is played back. Right (c): Countdown and image acquisition with flash.}
\label{fig:tutorial}
\end{figure}

In the interactive tutorial (IT) case, the process of positioning the phone is first demonstrated with an on-screen animation (Figure~\ref{fig:tutorial}). After viewing the complete animation, the user is guided by voice instructions to complete each of the steps in turn. When the camera is in the right position the user is instructed to `stop' and, after a countdown of three seconds, the camera flash light is activated to illuminate the eye evenly and the photo is taken. In the auditory guidance (AG) case, voice guidance is used for coarse positioning, e.g. `to the right', `up'. When the eye is recognised in the middle of the screen, the audio output consists of a sequence of short tones, which become more frequent with better camera positioning (cf. Geiger counter metaphor, ~\cite{holland2002audiogps}). As soon as the camera is in the correct position, the audio tone becomes continuous, the camera flash light is illuminated and the eye photo is automatically taken. In both IT and AG cases successful acquisition is indicated through a camera shutter sound and vibration feedback. At the end of the process, an image review screen is shown to the user with the option to retake should the iris be blurry.


\subsection{Implementation}
\textit{EyeGuide} is an Android app written in Java. To track the eye position we used the \texttt{openCV} library with the \texttt{Haar Feature-based Cascade Classifier for Object Detection}~\cite{haarCascade}, which is an AI-based model for image recognition tasks. To find the optimal parameters for the eye detection (\texttt{detectMultiScale} method, parameters \texttt{scaleFactor} and \texttt{neighbors}), we took 50 sample images, 28 with eyes and 22 without, with the rear camera of a Samsung Galaxy S7. The images without eyes showed other parts of the face around the eyes. We then analyzed the classifications (correct, false positive, false negative) for combinations of the parameters. Finally, we chose the parameters with the highest accuracy and speed. 

For the interactive tutorial (IT) case, graphics and animations were created using standard graphic design applications. To design the tutorial and the virtual agent with a focus on teaching motor skills, we adhered to the cognitive theory of multimedia learning~\cite{clark2016learning}, in particular the pre-training and segmenting principle. The voice instructions were recorded by author LD. To give feedback about the position of the eye in the auditory guidance (AG) case, the camera view was divided into five virtual rectangles; one in the center of the frame and four around the center. If an eye was recognised in one of the outer rectangles, speech guidance was given using the Android Text-To-Speech engine, e.g. ``to the right". The maximum frequency of the speech guidance was two seconds. When the eye was detected in the central rectangle, the guidance mechanism was changed to a sequence of short tones, which become more frequent with better camera positioning.

In both IT and AG cases, when the camera was optimally positioned, indicated either by `stop' (IT) or a continuous tone (AG), a full-resolution photo was automatically saved to the phone. In the capture phase, guidance was based only on position, as the available computing power and lower quality camera stream did not support analysing the sharpness of the image in real time.


\begin{figure}[!t]
\centering
  \includegraphics[width=\columnwidth]{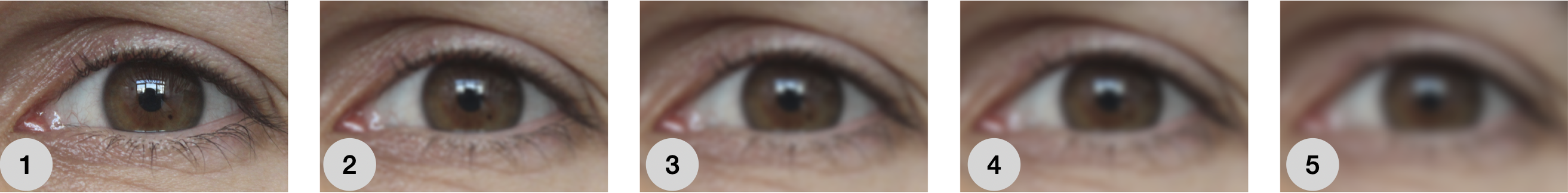}
  \caption{To rate the image sharpness in the \textit{EyeGuide} study, we used a series of images as a reference. From left to right: \textit{completely sharp (1), sharp (2), slightly out of focus (3), blurry (4), completely blurry (5)}.}~\label{fig:sharpness_rating}
\end{figure}

\begin{figure}[!hb]
\centering
  \includegraphics[width=\columnwidth]{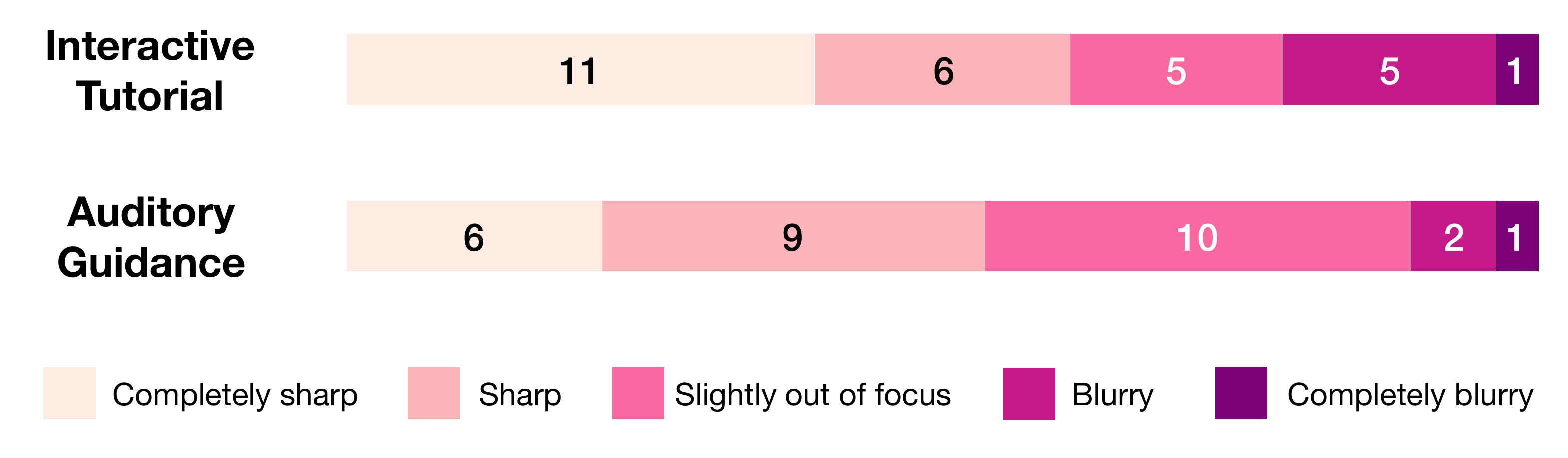}
  \caption{Distribution of image sharpness ratings. Each participant (n = 28) took two images, one for each eye. There were no significant differences in image quality (\textit{p} $>$ .05) between the two groups (IT and AG).}~\label{fig:imageQuality}
\end{figure}

\subsection{Evaluation}
To evaluate the user experience of \textit{EyeGuide} and its perceived diagnosis trustworthiness we conducted a user study. The study was conducted as a lab study and evaluated the two alternative designs, IT and AG, to enable comparison (Figure~\ref{fig:study_conditions}). We recruited 28 participants and assigned them to two groups (IT and AG) of 14. The IT group had a mean age of 23, while the AG group had an mean age of 22. Both groups consisted of six female and eight male participants. Five participants of the IT group and seven participants of the AG group wore glasses to correct their vision.

The study sessions were video recorded and all data collected during the study was stored anonymously. The participants were first introduced to the target of the study and its process and were then asked to sign a consent form. Afterwards, participants completed a demographic questionnaire and the TA-EG questionnaire~\cite{karrer2009technikaffinitat}, which measures technical affinity. Participants were encouraged to think-aloud during the test. The smartphone running the \textit{EyeGuide} app was handed to the participants and they proceeded to follow the instructions given by the app. At the end of the test, participants responded to open questions about the experience of using \textit{EyeGuide}.

\subsection{Results}
We firstly compared the number of attempts and time needed to take an acceptable eye image in both interactive tutorial (IT) and auditory guidance (AG) configurations. We conducted independent samples t-tests with $\alpha$ set to .05. Participants in the IT group (\textit{M$_{tries}$} = 1.8, \textit{SD} = .47) and in the AG group (\textit{M$_{tries}$} = 1.3, \textit{SD} = 1.19) did not differ in tries to capture the first eye, \textit{t}(26) = 1.46, \textit{p} $>$ .05). However, the duration needed to take photos of both left and right eyes was significantly lower, \textit{t}(26) = -2.17, \textit{p} $<$ .05, in the IT group (\textit{M$_{duration}$} = 151s, \textit{SD} = 115.3s) than in the AG group (\textit{M$_{duration}$} = 247s, \textit{SD} = 118s).


\begin{figure*}[b]
\begin{adjustbox}{max width=\textwidth,center}
\centering
\subfigure[\textit{No trust-building (baseline)}. Ishihara~\cite{ishihara1987test} plates were shown in all design alternatives, except for the assessment diagnosis mismatch.]{\label{fig:ishihara}\includegraphics[width=\width]{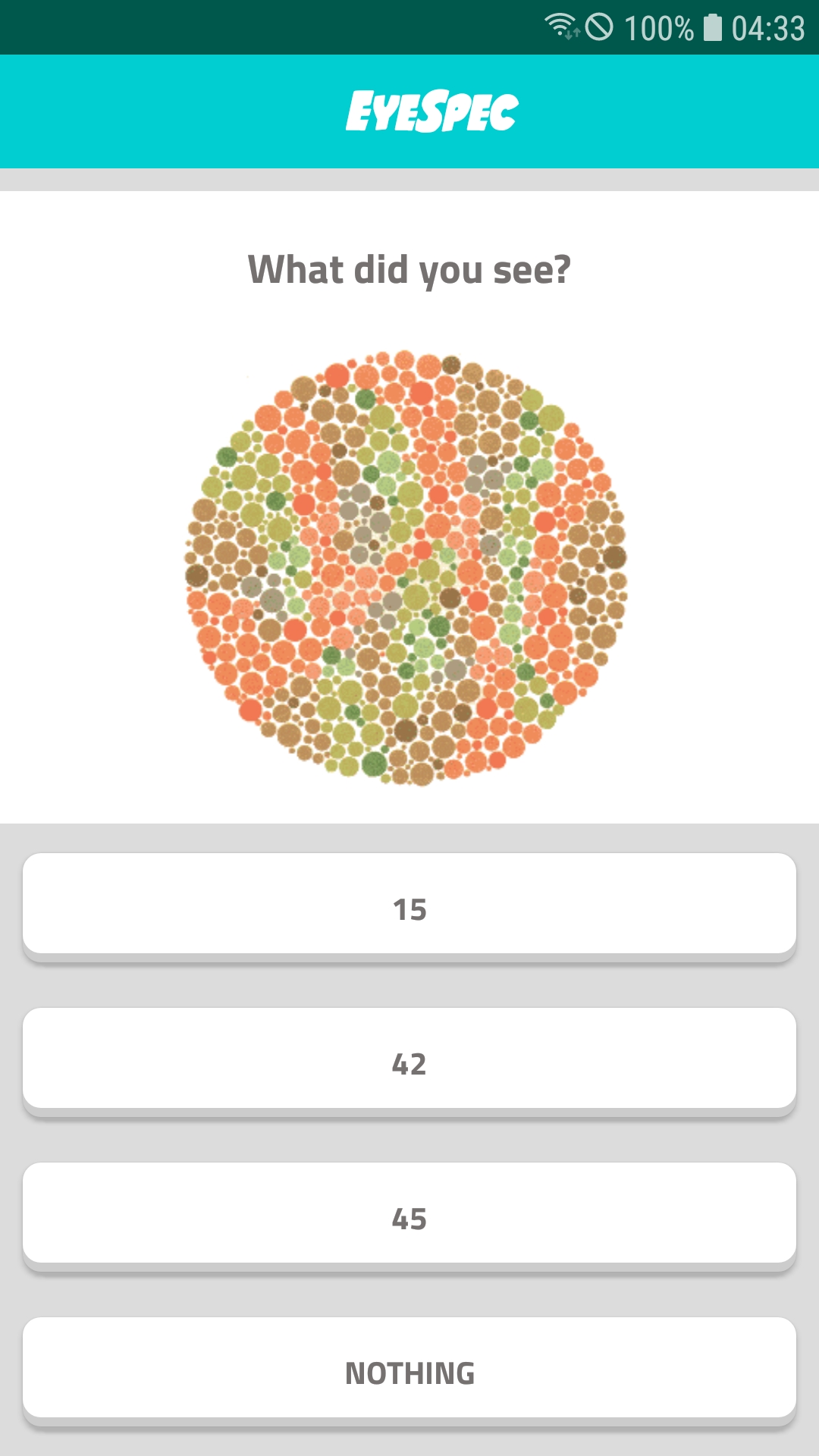}}\hspace{\horspace}
\subfigure[\textit{Assessment-diagnosis mismatch}. Participants provided speech samples that were not analyzed to allegedly detect color blindness.]{\label{fig:voice}\includegraphics[width=\width]{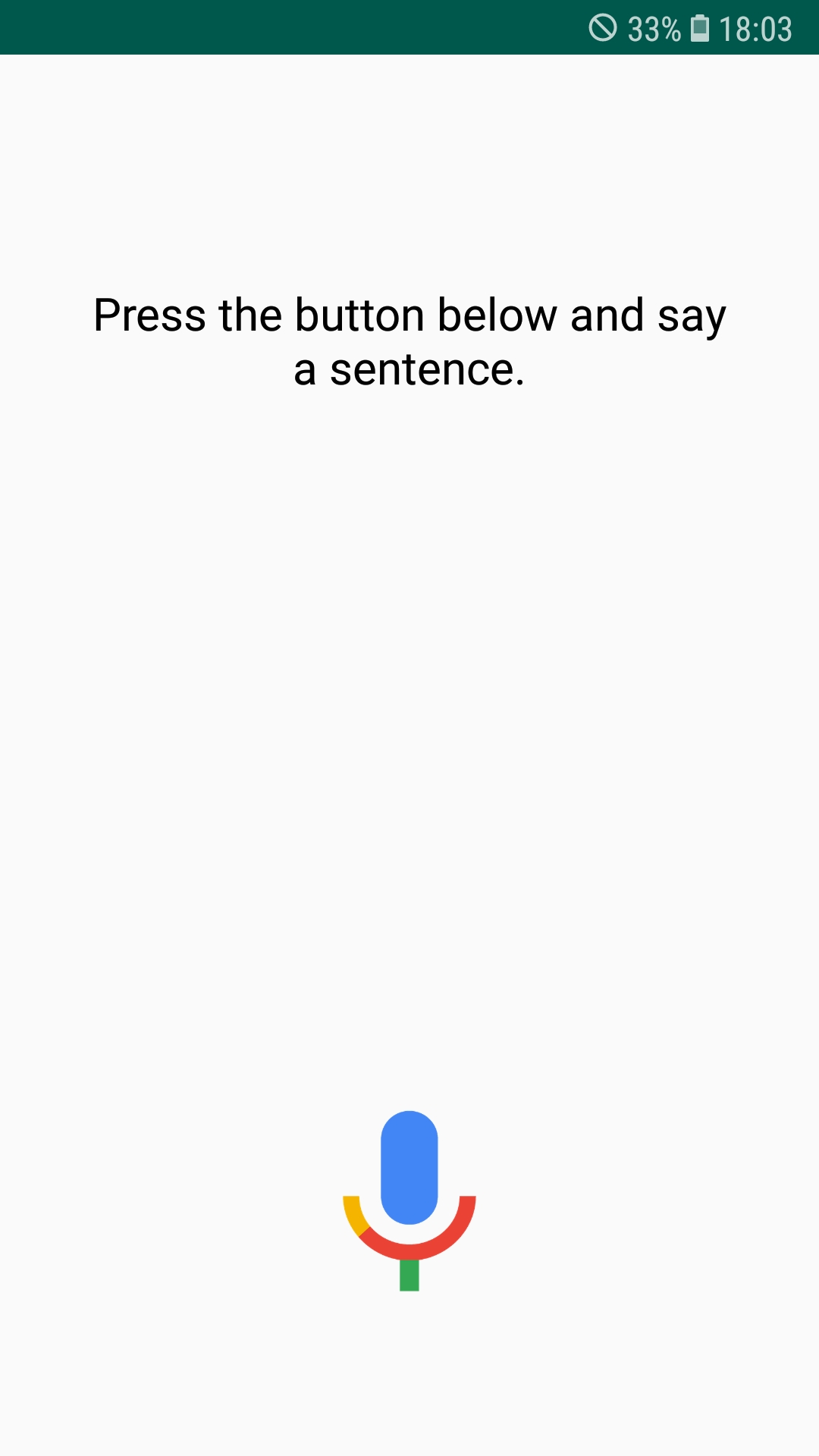}}\hspace{\horspace}
\subfigure[The \textit{disclaimer} explained that the assessment performed in the study did not replace an examination by a professional ophthalmologist.]{\label{fig:disclaimer}\includegraphics[width=\width]{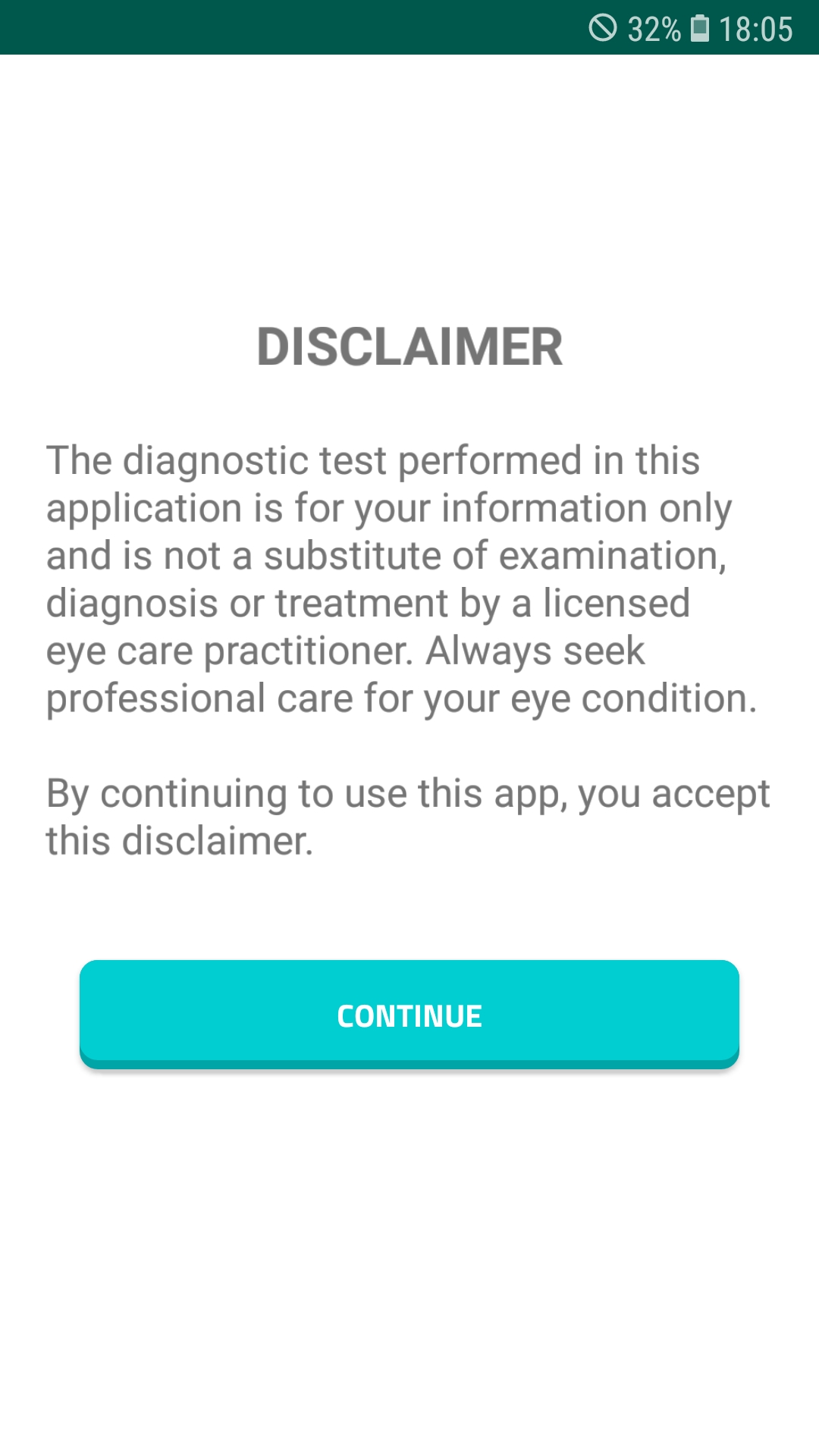}}\hspace{\horspace}
\subfigure[In the \textit{disease information} design variant, the participants were shown an explanation of possible causes for color blindness.]{\label{fig:disease_info}\includegraphics[width=\width]{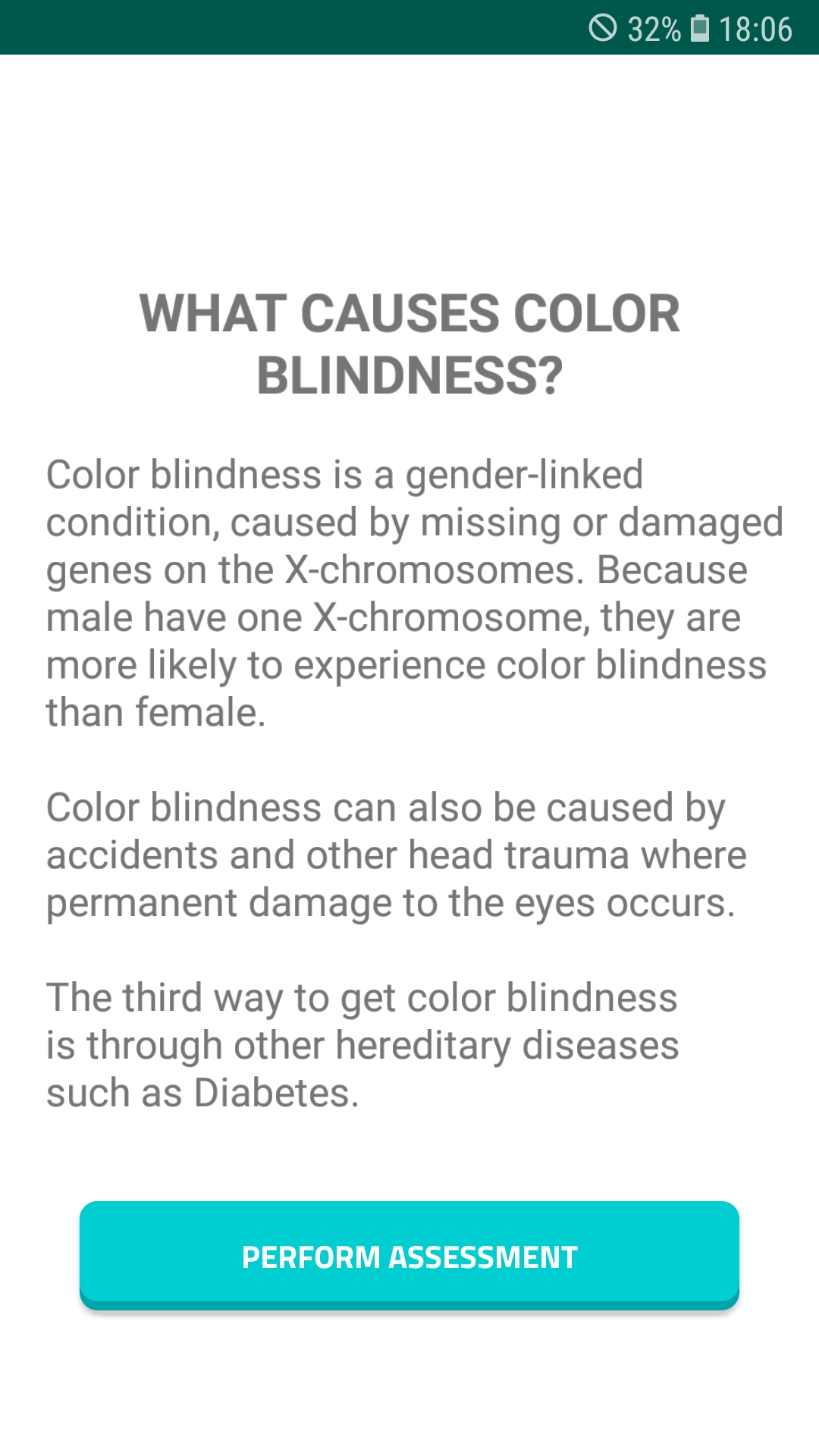}}\hspace{\horspace}
\end{adjustbox}
\caption{The four design alternatives of the trustworthiness study.}
\label{fig:tw_conditions}
\end{figure*}

Author LD rated the sharpness of the eye images by comparing them to a set of pre-defined reference images (Figure~\ref{fig:sharpness_rating}). The number of images rated with each quality level are shown in Figure~\ref{fig:imageQuality}. A majority of images was either completely sharp or sharp in both groups (IT: 17, AG: 15) and only one image was completely blurry in each of the groups. No significant difference between the two groups was apparent, \textit{t}(26) = .437, \textit{p} $>$ .05, with mean ratings for the IT group \textit{M$_{imageQuality}$} = 3.75, \textit{SD} = .87 and the AG group \textit{M$_{imageQuality}$} = 3.6, \textit{SD} = .86. Due to a technical error in the interactive tutorial condition, the image capture was triggered at a larger distance compared to the auditory guidance condition. This might have improved the image quality since a lower device-eye distance can cause blurry photos.



In the interactive tutorial (IT) group, the on-screen agent instructed the participant to watch the following steps before putting them into practice. This instruction was viewed and confirmed by every participant. However, as the tutorial started, six out of 14 participants tried to follow the instructions by turning the smartphone around while the first step of tutorial was shown. Two of the six realized that it was an instructive section of the tutorial when they were asked to press the `next' button, and turned the smartphone back around to complete watching the tutorial. The other four appeared to be confused on how to press the on-screen `next' button while physically following the tutorial steps. 
Eight of the 14 participants in the IT group had at least one occasion where the word `stop' was not played back, mostly due to to incorrect positioning of the device in front of the face. The participants then had to repeat the steps from the beginning. However, while the  smartphone was being moved back to the start position, sometimes one of the eyes was recognized, which triggered the `stop' instruction and resulting in participants capturing an image of their whole face. Other problems noted in the IT group included participants thinking they should capture both eyes at once, or not understanding that the photo would be taken automatically. Four out of five participants wearing glasses removed their glasses to take the photo even though they were not asked to do so. Participants were instructed to start taking each photo with a fully outstretched arm, however, after going through the process once, most participants did not fully outstretch when taking subsequent photos.

In the auditory guided (AG) group, participants first read the instructions before taking any photos. After turning the smartphone around so they faced the rear camera, six of the 14 participants felt unsure about the correct distance from the eye to hold the smartphone. Four of the participants were not sure if they were required to take the photo themselves or if the photo would be taken automatically. In addition, eight participants were not able to interpret the meaning of the  beeping sound. Two of the participants said that they do not know what to do when getting the voice instructions (to the left/right, up/down). Six participants turned the smartphone at least once to recheck the written instructions. In the AG group, four of the seven participants wearing glasses took their glasses off before taking the photos. 

To summarise, the time to take images in the interactive tutorial (IT) version of \textit{EyeGuide} was lower than in the audio guided (AG) version. The sequence of steps shown in the tutorial was clear to most study participants. However, we observed some usability issues, e.g. incorrectly carrying out physical instructions when participants had been instructed only to watch in the pre-training step.

\section{Trustworthiness Study}
Having evaluated different approaches for the tutorial and image capture phases, we also addressed the diagnosis feedback phase of smartphone-based eye disease detection. To gain insight into issues of trustworthiness in our targeted automated eye disease diagnosis app, we conducted a second experimental study. 

\subsection{Study Design}
\begin{figure}[!htbp]
\centering
  \includegraphics[width=.3\columnwidth]{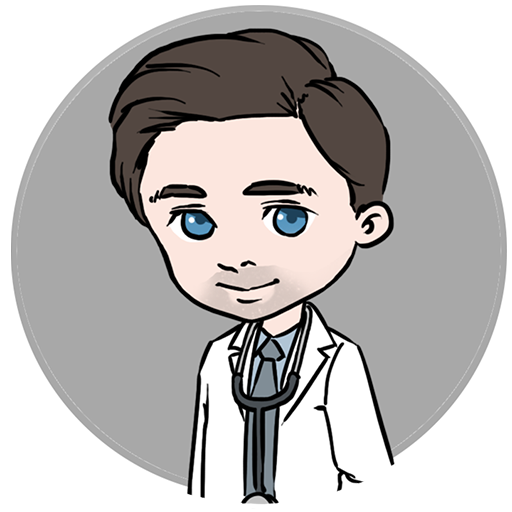}
  \caption{The virtual ophthalmologist avatar that leads through the color blindness test and diagnosis procedure.}~\label{fig:virtual_avatar}
\end{figure}

As the eye diseases we target with \textit{EyeGuide} are rare in healthy Western populations, for the purpose of our study we decided to simulate the detection of the more common ailment of color blindness. We did not conduct any actual diagnoses and chose color blindness as an arbitrary common condition that is related to vision. As a study probe, we developed a separate Android app that included four alternative design approaches (Figure \ref{fig:tw_conditions}) to building trust. Each of the alternatives is mapped to one or two dimensions of the trustworthiness model. The users' interaction with the app was guided by an animated avatar, that represented the virtual ophthalmologist (Figure \ref{fig:virtual_avatar}).  

The four alternative designs included in our application were:
\begin{enumerate}
\item \textit{No trust-building (baseline)}. This design represents the standard procedure for the assessment and diagnosis of color-blindness using the Ishihara test~\cite{ishihara1987test}.

\item \textit{Assessment-diagnosis mismatch}. This probe design examined if users trust the diagnosis even if there is a mismatch between the assessment and the diagnosis. Participants were asked to provide a speech sample to detect whether they have color blindness, even though there is no diagnostic relationship between these two modalities. This intervention is a negation to `Predictability' from the trustworthiness model~\cite{akter2011trustworthiness}. Predictability refers to the degree to which an application is expected to behave reliably by abiding to standard practices. 

\item \textit{Disclaimer}. With this design we aimed to find out whether informing the user that their data will not be used for wrong purposes and only with good intentions builds trust in the application. This targets the `Integrity' and `Benevolence' dimensions of the trustworthiness model. Benevolence refers to the good intentions of the application towards the user, whereas integrity refers to moral and ethical principles which an app should conform to. Therefore, to affirm both of these dimensions in the app, we display a disclaimer as well as terms and conditions at app start up.

\item \textit{Disease information} This design addressed the `Ability' dimension, which according to ~\cite{akter2011trustworthiness} is the most significant in the trustworthiness model. Ability refers to skills and competencies of the app that encourage the user to accept it, e.g. the application contains the desired knowledge about the task it performs. Addressing this, our design provided information in text form about color blindness to the user before actually starting the assessment.
\end{enumerate}

\subsection{Evaluation}
To compare the trustworthiness of our four design variants we carried out a user study. The study followed a between-subjects design with the independent variable \textit{trust-building feature (no trust-building, assessment-diagnosis mismatch, disclaimer, disease information)} and the dependent variable \textit{trustworthiness}. Participants were consecutively assigned to one of the four groups. Based on ~\citet{akter2011trustworthiness}'s trustworthiness model we developed the following hypotheses:

\begin{itemize}
    \item H1: Giving background information about the condition results in the highest trustworthiness score.
    
    \item H2: The lowest trustworthiness is observed when there is a mismatch between assessment and diagnosis. 

    \item H3: The highest benevolence score is achieved through providing a disclaimer as well as terms and conditions.

    \item H4: Integrity is highest through displaying a disclaimer and terms and conditions.

    \item H5: The ability score is highest when giving background information about the disease. 

\end{itemize}

\begin{figure*}
\centering
  \includegraphics[width=\textwidth]{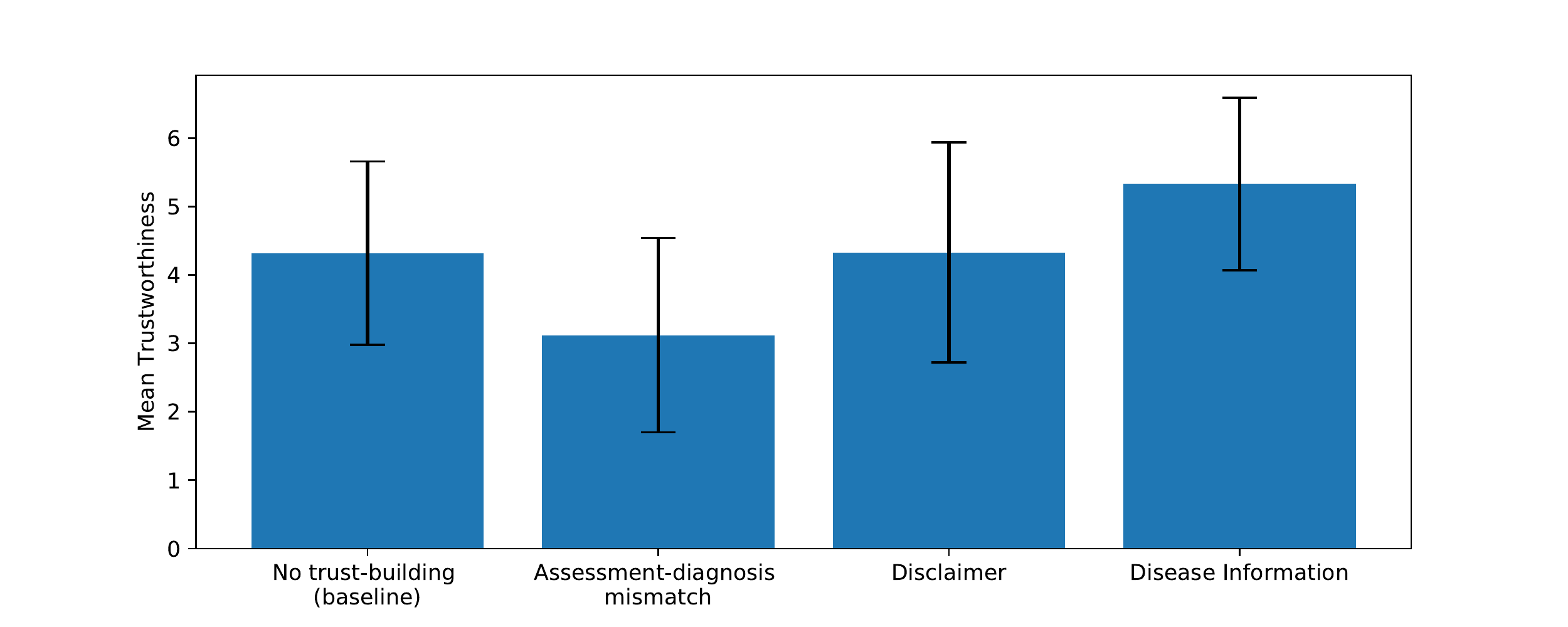}
  \caption{Trustworthiness scores between design alternatives. Error bars represent standard deviation.}~\label{fig:tw_scores}
\end{figure*}

At the start of the session, we informed participants about the study both verbally and in writing. Participants then completed a consent form and a demographic questionnaire. After that, participants were given a smartphone containing the study app, which they opened and followed the instructions given by the app. At the end of the study session participants filled out a trustworthiness questionnaire and were given an opportunity to provide open feedback on the test process. During the study, the experimenter took notes. The trustworthiness questionnaire developed for the study was a version of that developed by ~\cite{akter2011trustworthiness}, slightly modified for mHealth applications. 
The questionnaire consisted of multi-item scales with favorable psychometric properties. Each item in the questionnaire was measured in a structured arrangement on a 7-point Likert-type scale, ranging from 1 (strongly disagree) to 7 (strongly agree). There are 4 items for each of the dimensions of trustworthiness, and a final 4 questions directly investigating trust. 

We recruited 40 participants (\textit{M$_{Age}$} = 26.2, \textit{SD} = 6.0), 18 of which were female and 22 male. Most of the participants were students of the University of Bremen.

\subsection{Results}

We first report findings on the overall trustworthiness score followed by the results referring to its sub-dimensions. To test the effect of the independent variable design alternative on the dependent variable trustworthiness, we conducted a one-way ANOVA with the significance level $\alpha$ = .05. For further analysis of the differences between groups, we used planned contrasts. When referring to individual participants, we indicate the design alternative and the participant number within the condition, e.g. P1.2 is participant 2 in the no trust-building condition (no trust-building = 1, assessment-diagnosis mismatch = 2, disclaimer = 3, disease information = 4).

As expected, we observed differences in trustworthiness between the design alternatives (Figure \ref{fig:tw_scores}), \textit{F}(3,36) = 3.84, \textit{p} $<$ .05, partial \textit{$\eta^2$} = .24. When provided with disease information (\textit{M} = 5.33, \textit{SD} = 1.26), participants reported higher trustworthiness compared to the other design alternatives, \textit{t}(36) = 2.68, \textit{p} $<$ .05. Also, a mismatch between assessment and diagnosis led to the lowest trustworthiness, \textit{t}(36) = -2.86, \textit{p} $<$ .05.

While the integrity dimension score was affected by the design alternative, \textit{F}(3,36) = 5.23, \textit{p} $<$ .05, partial \textit{$\eta^2$} = .30, displaying a disclaimer did not lead to the highest integrity score, \textit{t}(36) = .79, \textit{p} $>$ .05. The descriptive data indicate that providing disease information (\textit{M} = 4.63, \textit{SD} = 1.29) might increase integrity.

Similarly, the design alternative had an effect on the benevolence score, \textit{F}(3,36) = 3.08, \textit{p} $<$ .05, partial \textit{$\eta^2$} = .20. However, contrary to our expectation, the disclaimer did not cause the highest benevolence, \textit{t}(36) = .54, \textit{p} $>$ .05. Again, there are descriptive indications that disease information (\textit{M} = 5.55, \textit{SD} = 1.49) is more effective in increasing benevolence.

Lastly, there was a difference in ability, targeted through disease information, between the design alternatives \textit{F}(3,36) = 6.42, \textit{p} $<$ .001, partial \textit{$\eta^2$} = .35. Consistent with our hypothesis, information about color-blindness led to higher ability, \textit{t}(36) = 3.18, \textit{p} $<$ .05.


In the \textit{no trust-building} group, seven out of ten participants complimented the app with statements like ``nicely done" (P 2.10) and ``fast" (2.9). When questioned whether they would trust the app or not only two of the participants said yes. Participant 2.2 said she might have trusted it if she didn't know already that she is not color blind. Three out of ten participants in the \textit{assessment-diagnosis mismatch} group were confused when presented with the results. Six participants questioned the diagnosis of color blindness based on speech samples. When shown the \textit{disclaimer and terms and conditions}, seven out of ten participants read both screens. One individual skipped the terms and conditions text. All of the participants performed the Ishihara test correctly. One of the participants mentioned that they liked the app, especially the dialogue with the virtual ophthalmologist (P 3.6). Participant 3.5 commented that the app seems to have good intentions. On questioning them about their trust in the app, only participant 3.5 said that she trusted it, whilst  participant 3.6 said she might trust it, but was a little hesitant. 

We observed that nine out of ten participants in the \textit{disease information} group read both the information screens, i.e. ``what is color blindness?" and ``what causes color blindness?". Participant 4.2 skipped the ``what is color blindness?" screen. 
Six participants agreed that the app contains knowledge about the disease. Three participants rated the app with comments such as ``user-friendly" (P 4.4) and ``nice user interface" (4.10). Five out of ten participants said that they trusted the app. Participant 4.2 said she would trust the app if it was used in cooperation with a doctor.




\section{Discussion}
Through two studies we firstly examined the usability of \textit{EyeGuide}, a mHealth app to take medical eye photos and, secondly, identified design considerations to ensure trustworthiness of mHealth apps in general. While the evaluation of \textit{EyeGuide} addressed the image capture phase of an automated eye disease diagnosis app, the trustworthiness study informed on design approaches to increase trustworthiness in the diagnosis feedback phase.

\subsection{EyeGuide}
As some participants did not realise that the photo would be taken automatically and that they should only capture one eye at a time, not all tutorial instructions were understood correctly. Similar confusions about the meaning of the voice instructions (to the left/right, up/down) may have been due to a misunderstanding if the object to be moved was the smartphone or their head.  
On some occasions, participants repeated steps or activated functions by mistake because they were holding the smartphone in an incorrect position. Participants in both groups were confused when they were required to confirm the taken image. 

Participants that wore glasses had similar problems with both concepts.  All the photos taken by participants wearing glasses were of poor quality. Particularly, all those rated as completely out of focus were taken by glasses wearers and other images captured by glasses wearers were either blurry or slightly out of focus. To remedy this, participants removed their glasses - in hindsight this should have been instructed by the app.


A possible reason for the interactive tutorial group being faster in taking the eye images than the auditory guided tutorial group is that the auditory guided tutorial group took longer to find the correct device position. In contrast, the interactive tutorial group just had to follow the given instructions. Additionally, auditory guided participants often turned the smartphone around to check the instructions again, which added additional time. The mean time difference between the IT and AG group to complete the two images was on average about 100 seconds. In a real-world scenario, this has probably implications on the frustration levels of users, potentially leading to drop-outs in the AG concept.


In the interactive tutorial, some participants had trouble distinguishing between the sections of the tutorial where they were supposed to only watch the instructions and the sections where they had to actively carry out the instructions, e.g. actually turn around the phone. While the agent instructed the participants to first watch the steps before following them together, this was apparently not salient enough. We suggest a multi-modal indication of whether to watch only or carry out the instructions, e.g. a voice and text instruction.

In the auditory guidance group, many users did not recognize the Geiger counter metaphor, i.e. a beep tone with shorter intervals as they approached the target position.

Overall, we believe that the interactive tutorial concept had a better user experience. While some errors were detrimental to task completion and image quality, the concept of providing instructions that have to be carried out afterwards has shown to be effective in the context of taking eye images.

\subsection{Trustworthiness}
Our results indicate that providing users with disease-specific information increases the trustworthiness and potentially also positively impacts benevolence and integrity. This is consistent with ~\citet{akter2011trustworthiness} who regarded the ability dimension, i.e. the perceived skills and competences of an app that encourage users to use it, as the most significant. Therefore, we confirm our hypothesis H1. On the other hand, trustworthiness is lowest when the assessment and the diagnosis modalities do not match, we can confirm H2. The disclaimer did not lead to the highest benevolence and integrity dimension, meaning we reject our hypotheses H3 and H4. There are descriptive trends that suggest that disease information is the most effective factor to increase these two trustworthiness dimensions. Furthermore, participants reported the highest trustworthiness when shown disease information; we therefore confirm H5.

\subsection{Diagnostic Limitations}
Whilst we propose that \textit{EyeGuide} requires less expert knowledge and provides better accessibility compared to existing smartphone adaptors for retinal imaging, it is important to note that the spectrum of detectable diseases between the two approaches differs. Through fundus images taken using smartphone adaptors, a range of diseases such as diabetic retinopathy, glaucoma, and age-related macular degeneration can be diagnosed ~\cite{Roberts1999}; 
with these images, algorithmic approaches can be use to predict cardiovascular risk factors~\cite{poplin2018prediction}. Smartphone cameras without additional lens adaptors do not allow for retina inspection and can only identify symptoms of diseases that are visible on the outside of the eye (e.g., conjunctiva or cornea). However, as some of the diseases visible with the naked eye, e.g. cataracts, are preventable~\cite{WHO2013}, there is an obvious need for better screening, education and intervention in eye care. We argue that easy access to \textit{EyeGuide} through a smartphone and the possibility of detecting cataracts, which are responsible for 33\% of global blindness, with smartphone images justifies further research on this topic.


\subsection{Contribution and Future Work}
Given the recent development of AI-based diagnosis approaches, we argue that there are a large amount of potential application areas for \textit{EyeGuide} to be integrated as part of existing eye-disease screening and diagnosis tools. The interactive tutorial, based on principles of multimedia learning, was successful in preparing most users to capture high quality images. Minor usability issues were identified in the user study and will be addressed in the following design iterations of \textit{EyeGuide}.

We are one of the first to explore solutions for guiding image capture for eye images. While similar approaches in the domain of dermatology~\cite{Skinvision} include features to automatically trigger image capture, they do not provide any support for device positioning. Other authors~\cite{lo2013yes} described the use of verbal guidance for taking portrait photos, however, our interactive tutorial (IT) approach combines a tutorial and verbal guidance as an integrated solution.

Based on our findings, we suggest that the designers of mHealth applications should include disease-specific information to ensure trustworthiness. 
While we focused on color blindness in our trustworthiness study, our findings are generalizable to other mHealth fields beyond eye-related diseases. As future work we plan to extend our study to encompass a broader sample, e.g. the elderly, and to explore the use of our \textit{EyeGuide} approach as part of iris imaging for biometric authentication~\cite{Thavalengal2015Iris,Raja:2014,raja2015multi}.

\section{Conclusion}
\balance
In this paper, we presented two studies exploring the usability and trustworthiness of eye-related mHealth applications. An interactive and multi-modal tutorial that demonstrates correct device positioning was successful in reducing acquisition time. An easy to use, self-administered app to support screening and diagnosis of eye-related conditions can increase eye care access, especially in developing countries. Providing disease-specific background information was shown to be the most effective intervention to increase trustworthiness in the diagnosis. Our findings are relevant to the designers and developers of mHealth applications in the area of ophthalmology and other medical fields.

\section*{Acknowledgments}

We would like to thank all participants for contributing to our studies. This research was supported by a Lichtenberg Professorship from the Volkswagen Foundation, the BMWi funded network KI-SIGS (grant 01MK20012) and the Leibniz ScienceCampus Bremen Digital Public Health, which is jointly funded by the Leibniz Association (W4/2018), the Federal State of Bremen and the Leibniz Institute for Prevention Research and Epidemiology – BIPS.





\bibliographystyle{IEEEtranN}
\bibliography{IEEEabrv,references}


\end{document}